# Three-dimensional Coherent X-ray Diffraction Imaging via Deep Convolutional Neural Networks

Longlong Wu[1,2]*, Shinjae Yoo[1], Ana F. Suzana[2], Tadesse A. Assefa[2,3], Jiecheng Diao[4], Ross J. Harder[5], Wonsuk Cha[5] and Ian K. Robinson[2, 4†]

[1]*Computational Science Initiative, Brookhaven National Laboratory, Upton, NY 11973, USA*

[2]*Condensed Matter Physics and Materials Science Department, Brookhaven National Laboratory, Upton, NY 11973, USA*

[3]*Stanford Institute for Materials and Energy Sciences, SLAC National Accelerator Laboratory, Menlo Park, California 94025, USA*

[4]*London Centre for Nanotechnology, University College London, London, WC1E 6BT, United Kingdom.*

[5]*Advanced Photon Source, Argonne, Illinois 60439, USA*

*Correspondence authors' e-mail: *lwu@bnl.gov, †irobinson@bnl.gov*

## ABSTRACT

As a critical component of coherent X-ray diffraction imaging (CDI), phase retrieval has been extensively applied in X-ray structural science to recover the 3D morphological information inside measured particles. Despite meeting all the oversampling requirements of Sayre and Shannon, current phase retrieval approaches still have trouble achieving a unique inversion of experimental data in the presence of noise. Here, we propose to overcome this limitation by incorporating a 3D Machine Learning (ML) model combining (optional) supervised learning with transfer learning. The trained ML model can rapidly provide an immediate result with high accuracy which could benefit real-time experiments, and the predicted result can be further refined with transfer learning. More significantly, the proposed ML model can be used without any prior training to learn the missing phases of an image based on minimization of an appropriate 'loss function' alone. We demonstrate significantly improved performance with experimental Bragg CDI data over traditional iterative phase retrieval algorithms.

**INTRODUCTION**

Coherent X-ray diffraction imaging (CDI) has been widely utilized to characterize the internal three-dimensional (3D) structure of single particles[1-4]. Particularly, Bragg CDI has emerged as a promising technique for 3D strain imaging of crystalline particles[5-11]. As modern X-ray sources, such as diffraction-limited storage rings and fourth-generation X-ray free-electron lasers, are developing worldwide to provide higher coherent flux densities, time-resolved and *in-situ* CDI experiments for single-particle imaging are becoming more and more capable to explore small particles' dynamical phenomena such as driven melting, thermal fluctuation, driven phase transitions, catalysis, and high-pressure phenomena[12-17]. Due to the lost phase information in measured coherent X-ray diffraction signals, it is necessary to use phase retrieval[18-21] as a key component of CDI, to reconstruct the real-space 3D images with morphological details from the measured signals.

Until now, the extensively used approach for CDI phase retrieval is the iterative methods, such as the hybrid input-output (HIO) method by Fienup[19], the difference map (DM) by Elser[22], and the relaxed averaged alternating reflection (RAAR) method by Luke[23]. In general, these iterative phase retrieval methods can be expressed as successive projections[22,24]. Theoretically, for a finite object, when the modulus of its Fourier Transform is well oversampled, a unique solution is guaranteed for these methods[19,20]. However, for experimental data with inherent noise, these projection-based methods are found to struggle with local minima, which leads to an ambiguous, rather than unique, solution[25]. Thus, when inverting coherent X-ray diffraction patterns, conventional iterative methods typically need thousands of iterations and switch algorithms to confidently converge to a reproducible solution and require tuning of many algorithmic parameters and expert strategies[24,26,27]. Because these methods are based on projections, the calculated error usually is only used to monitor the convergence and rarely used as feedback to adjust the related algorithmic parameters, which makes these methods sensitive to their initialization conditions.

For phase retrieval, deep-neural-network-based ML methods have recently shown a significant advantage in providing rapid reconstruction results in a CDI experiment[28-30]. There has been rapid progress for 2-dimensional (2D) phase retrieval using convolutional neural networks (CNN) recently[28,30]. Meanwhile, an adaptive ML-based approach for 3D phase retrieval has been demonstrated by using spherical harmonics[29]. However, so far, most of the proposed neural networks employ a supervised learning approach, matching input diffraction patterns in reciprocal space to output particle morphological information in real space, which usually needs large training datasets to train the neural network so that it can represent a universal approximation function. When a deep neural network is trained with limited data, its ability to generalize as a universal function is reduced, as seen in the accuracy of the reconstructed results, and a subsequent refinement procedure is needed to follow the supervised learning[28,29]. Furthermore, in practice, it is difficult to obtain enough ground-truth experimental coherent X-ray diffraction data for training. When applied to experimental data, an ML model trained with less data than ideal may also suffer and may need very lengthy experiment-specific retraining.

Here, we demonstrate a comprehensive 3D CNN-based approach to reconstruct the interior complex morphological information of a range of nanoparticles from their measured coherent X-ray diffraction patterns. When trained in a supervised learning approach, this CNN model can be applied to real-time 3D single-particle imaging experiments, for example, using an X-ray Free-electron Laser (XFEL)[31]. Further, while seeking to improve the accuracy of the result, we find that the trained CNN model can also be used in an unsupervised transfer learning mode. We demonstrate significantly improved accuracy with both simulated data as well as experimental data.

Additionally, when recovering the 3D structure of a particle with the unsupervised learning approach, we find no distinguishable difference between the quality of the obtained results whether the pre-trained ML model is used (*i.e.,* transfer learning) or the randomly initialized model is used, except for their convergence speed. This is important in situations where the acquisition of training data is challenging. With the unsupervised learning approach, the flexibility of the self-defined loss function in the CNN model makes this method more robust to coherent X-ray diffraction data of lower quality than the traditional approach.

## RESULTS AND DISCUSSION

## ML model and data sets

Generally, for coherent X-ray diffraction imaging experiments, either forward-scattering CDI or for Bragg CDI[32,33], the measured X-ray diffraction intensity $I(\mathbf{Q})$ is proportional to the modulus squared of the Fourier Transform of a complex field $\rho(\mathbf{r})$:

$$I(\mathbf{Q}) = \left| \int \rho(\mathbf{r}) e^{i\mathbf{Q}\cdot\mathbf{r}} \, d\mathbf{r} \right|^2,$$ (1)

where $\mathbf{Q} = \mathbf{q} - \mathbf{h}$, and $\mathbf{q} = \mathbf{k}_f - \mathbf{k}_i$ is the momentum transfer defined by the incident and diffracted X-ray wavevectors $\mathbf{k}_i$ and $\mathbf{k}_f$. Here, $\mathbf{h}$ equals to zero for a forward CDI experiment, and $\mathbf{h}$ is a reciprocal lattice vector of the crystal in a Bragg CDI experiment. In equation (1), the complex field $\rho(\mathbf{r})$ is related to the local complex refractive index of a particle in a forward CDI experiment and in Bragg diffraction geometry, it mainly represents the local crystal lattice strain inside a particle[7,13]. In all cases, this complex-valued structure information inside the particle could also be expressed as $\rho(\mathbf{r}) = s(\mathbf{r}) e^{i\phi(\mathbf{r})}$, where $s(\mathbf{r})$ and $\phi(\mathbf{r})$ are the corresponding amplitude and phase distributions of the measured particle, separately.

The goal of a CDI experiment is to numerically obtain this complex particle density function uniquely in real space, whose modulus squared of the Fourier Transform best matches the experimental coherent X-ray diffraction intensity distribution of the measured particle[20]. As shown in Fig.1, our developed deep neural network for 3D coherent X-ray diffraction imaging adopts the typical 'encoder-decoder' architecture. It takes the amplitude of the 3D coherent X-ray diffraction pattern in reciprocal space as input and outputs the real-space amplitude and phase images. As presented in Fig. 1, the proposed model is implemented using an architecture composed entirely of 3D convolutional, max-pooling, and upsampling layers. The model adopts the general convolutional encoder-decoder network architecture, which has three main parts: a 3D convolutional encoder that encodes the input x-ray diffraction data through a series of convolutional blocks, followed by two decoder parts which utilize the encoded result to generate the real-space amplitude and phase information of the measured particles. In this 3D CNN, the leaky rectified linear unit (LRLU)[34] is used for all activation functions except for the final 3D convolutional layer, where the rectified linear unit (RLU) activation function is used. The modules used in Fig. 1 to connect the input from the previous layer to the next layer's output are convolution blocks (3×3×3 convolution + LRLU + BN, where BN refers to batch normalization), followed by convolution blocks (3×1×1 convolution + 1×3×1 convolution + 1×1×3 convolution + LRLU + BN). It should also be mentioned that the array size of output particle image arrays (*i.e.,* amplitude and phase) in each dimension is half of the size of the input diffraction data to keep the problem overdetermined.

**Supervised learning approach**

In the deep neural network supervised learning method, the quantity and diversity of the training dataset directly affect the network's performance when unknown data are presented. In the real world, the complex structure $\rho(\mathbf{r}) = s(\mathbf{r})e^{i\phi(\mathbf{r})}$ for a particle varies a lot from particle to particle. For demonstration purposes, a shape known as a superellipsoid is used to describe the particle shape $s(\mathbf{r})$ and a 3D Gaussian correlated profile is used to describe the corresponding phase $\phi(\mathbf{r})$ distribution (see Methods for details). Then, after the generated particle is randomly rotated in real space, a 3D coherent diffraction pattern is obtained by Fourier Transformation. Only the amplitude information of the diffraction pattern is kept for training, and the phase information is discarded.

By applying this method with a wide range of random parameters, we simulated 30,000 3D diffraction patterns and used them to train the CNN model. With the corresponding particles known *a priori*, the proposed 3D CNN model was trained in a supervised learning approach, by solving

$$l_{\text{s}} = \arg\min_{\rho_{\text{p}}}\left[\rho_{\text{p}}(\mathbf{r}) - \rho_{\text{g}}(\mathbf{r})\right], \tag{2}$$

where, $\rho_{\text{p}}$ is the output from the CNN model, and $\rho_{\text{g}}$ is the corresponding ground truth for the complex particle. For this loss function $l_{\text{s}}$, which was minimized during the training, we used a combination of the relative root mean square error $\chi$, and the modified Pearson correlation coefficient $r_{\text{p}}$ (see Methods for details) to measure the agreement between the output amplitude and phase images of the predicted particles with their ground truth both in real and reciprocal space. This is appropriate for diffraction data with a large dynamic range. The $\chi$ is dominated by the strong central part of the diffraction pattern. The $r_{\text{p}}$ is a statistical metric that measures the linear similarity between two variables[35]. When training the 3D CNN model, the prepared training data were divided into two disjoint sets, where 95% of them were used to train the model, and the rest of them were used for validation.

While the 3D CNN model was being trained by the supervised learning approach, Supplementary Figure 1 shows the training and validation loss as a function of the training epochs. It can be seen that the loss for the validation testing is generally continually decreasing. After 100 training epochs, the loss for the validation can reach 0.031, which illustrates that the proposed 3D CNN model can already provide a highly accurate estimation of the reconstruction. The nearly identical losses for training and validation result from an early stop placed on the training to avoid overfitting. To demonstrate the performance of this trained CNN model, Fig. 2 shows four representative predictions from test diffraction patterns, not used for training the CNN model. The predicted amplitude and phase of the particles show excellent agreement with their ground truth. This CNN model is an ML method of phase retrieval, which provides a very fast inversion of a diffraction pattern (~9 ms computation in our hardware). Unlike an iterative phase-retrieval method, this could be very useful in a real-time 3D CDI experiment, for example to capture movies of a moving or evolving object.

When the CNN model learns to match input coherent diffraction data to output particle data, it does not only learn to solve the data fitting problem but also incorporates comprehensive prior information (such as support size or phase range of a particle) in a data-driven manner[36]. Perhaps the greatest strength for the ML-based phase retrieval method is that the model can learn far more complicated prior information. The ability to ultimately learn both the best possible inverse solver and the specific prior information makes the model very powerful.

**Unsupervised learning approach**

Since the supervised ML-based approach is data-driven, sometimes the predicted results might miss subtle features in the data which were not captured by the training. To improve the quality of the obtained reconstruction, we have developed a refinement procedure by using an unsupervised transfer learning approach. This refinement improves the reconstruction of a single diffraction pattern at a time. The problem of phase retrieval for coherent X-ray imaging experiments can also be considered to be an optimization problem[22,37], expressed as

$$l_\mathrm{u} = \arg\min_{\rho_\mathrm{p}} \left[ \left| FT\rho_\mathrm{p}(\mathbf{r}) \right|^2 - I_\mathrm{m}(\mathbf{Q}) \right], \tag{3}$$

where, $l_\mathrm{u}$ is the loss function for unsupervised learning, which describes the difference between the numerically obtained particle $\rho_\mathrm{p}(\mathbf{r})$ and the measured coherent X-ray diffraction intensity $I_\mathrm{m}(\mathbf{Q})$. $FT$ represents the Fourier Transform operation.

In Fig. 3, we demonstrate that the proposed unsupervised transfer learning approach can further improve the reconstruction quality and reach a high accuracy, comparable with the best iterative algorithms. As shown in Fig. 3(a), we demonstrate this approach with a 3D diffraction pattern, which is obtained with the parameters that are different from the parameters used to generate the training data. The corresponding real-space particle is given in Fig. 3b. As can be seen from Fig. 3a and 3b, while the particle shape is symmetrical, the broken symmetry of the diffraction pattern results from the internal asymmetry of the real-space 3D phase distribution, which is common in Bragg CDI from particles with strain distributions. The trained 3D CNN model yields the reconstructed amplitude and phase structure shown in Fig. 3c, with a corresponding estimated error of 0.13. Compared with its ground truth in Fig. 3b, the trained CNN model gives a relatively poor prediction, indicating that a refinement is necessary, because features of the input diffraction are not fully captured by the training. Here, the pre-trained CNN model from the supervised learning was then used in the unsupervised transfer learning to further refine the reconstructed result using the loss function defined in Eq. (3) (see Methods for details). Figure 3e shows the result of this unsupervised transfer learning approach, and Figure 3f shows the trend of the corresponding loss (or error metric) with training epoch. After this refinement approach on a noise-free diffraction pattern of a test particle, significant improvement was achieved, where the error decreased from 0.13 to $2\times10^{-6}$. The time cost for the transfer learning is ~28.67 ms per epoch (*i.e.,* ~3.19 h for $4\times10^4$ epochs) in our hardware.

When using the unsupervised transfer learning approach, to explore the importance of the pre-training for this CNN model, we further tested the model with the same coherent x-ray diffraction data by using randomly initialized bias and weight parameters (*i.e.,* without transfer). Figure 3g and 3h show the corresponding obtained diffraction pattern and particle by using this method separately, with the corresponding loss given in Fig. 3f. Comparing Figs. 3(g-h) with Figs. 3(d-e), it can be seen that there is no obvious difference between the final reconstructed results. This is a significant discovery: the ability of the Neural Network to retrieve phases directly without pre-training. As presented, Figure 3f shows that the loss converges faster for the CNN model after pre-training than the model with random initialization, however there is no significant difference between the final reconstructed results. To further quantify this effect on the final results, Figure 3i presents the Fourier Spectral Weight (FSW) of the reconstructed results from the two different methods, which is obtained by integrating the reconstructed diffraction amplitude of the predicted particle over shells of constant $|\mathbf{Q}|$. As shown in Fig. 3i, there is no noticeable difference between the FSWs from the two predicted results, which indicates that the two methods agree quite well at all spatial frequencies.

**Performance of 3D CNN model on experimental data.**

Since the internal structure of a crystalline particle is usually unknown in CDI experiments, it is vital that our proposed ML approach gives a credible reconstruction result for phase retrieval in the presence of unavoidable noise. With CDI experiments, there is little prior knowledge of the structure available for building a training dataset. Fig. 4(a-d) shows isosurface renditions of four very different experimental Bragg coherent X-ray diffraction patterns of individual $SrTiO_3$, $BaTiO_3$, Pd, and Au nanocrystals (see Methods for details), which were measured at beamline 34-ID-C of the Advanced Photon Source using methods reported by Robinson & Harder[7]. From Fig. 4(a-d), these four 3D Bragg coherent X-ray different patterns have different diffraction fringe spacings and directions, indicating their distinct sizes and facets in real space.

By using our 3D CNN model with the proposed unsupervised transfer learning approach, the corresponding predicted results are shown in Fig. 4(e-l). Fig. 4(e-h) shows the corresponding calculated X-ray diffraction intensities, obtained as the modulus squared of the Fourier Transform of the predicted CNN model structures shown in Fig. 4(i-l). There is excellent agreement between the experimental and calculated X-ray diffraction patterns, confirming that the CNN model can reach a high reconstruction accuracy. This result also extends to the randomly initialized CNN model with experimental data, for which Supplementary Figure 2 shows the corresponding reconstructed results with the same four sets of experimental Bragg CDI data. Comparing Fig. 4 and Supplementary Figure 2, it can be seen that there are no obvious differences between the final reconstructed particles, which endorses the capability of the untrained CNN model in the presence of experimental noise.

Given the tendency of the conventional iterative algorithms to show imperfect convergence and non-degenerate solutions with real (noisy) experimental data, as mentioned in the introduction, we investigated the reproducibility of the solutions obtained with the untrained CNN model by using different random numbers seeding the calculation. Using the coherent X-ray diffraction pattern in Fig. 4(a), we repeated the reconstruction 100 times with two different methods: the CNN model with random parameters (*i.e.,* untrained model) in the unsupervised learning approach and the conventional iteration method with random initialization (see Methods for details and Supplementary Figure 3 for the same comparison using simulated data). As shown in Fig. 5, the corresponding statistical error was found to have multiple solutions in both cases with roughly the same $\chi^2 = 0.0241 \pm 0.0005$ (standard deviation). The $r_p$ error of 0.9922 for ML was better than 0.9915 for iterative. The $r_p$ is statistical metric that measures the similarity between two variables. However, $\chi^2$ is usually dominated by the strong central part of a diffraction pattern. Thus, this difference is probably because the calculated loss (or error) explicitly optimized the $r_p$ and $\chi^2$ together. It is reassuring that the conventional iterative method performs so well since it is based only on projection, although it shows a slightly wider distribution of solutions. We also noticed qualitatively that the solutions from the ML method appeared sharper and had flatter, better-distinguished facets than the iterative algorithms, as can be seen in Fig. 6 and Supplementary Figure 4.

Based on above results, it can be concluded that our 3D CNN approach has great potential to be applied to asymmetric data previously untested owing to the need to solve for a complex density function whether there is enough training dataset to obtain a well-optimized ML learning model or not. When there are enough suitable data to train the CNN model, it can be used in a real-time experiment, such as in a single-shot XFEL setup, to provide a rapid estimation of the reconstruction. When needed, a subsequent unsupervised transfer learning refinement can then make the final predicted results reach the possible maximum accuracy.

Furthermore, the unsupervised learning approach makes it possible to use an untrained ML model for ab-initio phase retrieval of the structure of sub-micron-sized particles in 3D. This will be valuable when building a related training dataset for the training of an ML model is challenging. Additionally, in the unsupervised learning approach, the self-defined loss function, used for feedback to optimize the prediction, makes the ML model more powerful than the conventional iterative methods, where the calculated error during iteration is only used to monitor the convergence. In this work, we used a combination of Correlation Coefficient and $\chi^2$ error to define the loss function for the unsupervised learning, which extracted more sensitivity to the weaker features in the data than the $\chi^2$ error alone. However, one is not limited to this form and could further extend it for different phase-retrieval problems, for example using likelihood function to account for the statistical error.

In conclusion, we have demonstrated a comprehensive ML approach for the 3D reconstruction of single-particle structures in real space from their experimental coherent X-ray diffraction intensities in reciprocal space. The trained CNN model can provide immediate high accuracy results, which will benefit a real-time CDI experiment. More importantly, we found that the unsupervised learning approach was able to reach a high reconstruction accuracy, comparable with traditional methods, either starting from a pre-trained model (*i.e.,* transfer learning) or just a purely random configuration. The flexibility of the self-defined loss function in the ML model should make the CNN model more robust to experimental coherent diffraction data when used in unsupervised learning. The quality of the images obtained in the four examples shown here is better than can be achieved with current state-of-the-art iterative algorithms in use today. We believe our results will see very broad applications in coherent x-ray diffraction imaging and related research fields. This will also have significant effect on neural network design, where the combination of supervised and unsupervised learning can be generalized to solve other phase retrieval problems.

**METHODS**

**3D Training Dataset**

The 3D diffraction intensities were generated by taking the Fourier Transform of the simulated complex-valued 3D particles $\rho(\mathbf{r}) = s(\mathbf{r})e^{i\phi(\mathbf{r})}$, created from the particle's amplitudes $s(\mathbf{r})$ and phases $\phi(\mathbf{r})$. The simulated particles $\rho(\mathbf{r})$ were then randomly rotated in 3D. Only the amplitude of the computed 3D diffraction intensities was retained for both training and testing of the CNN model. We used a superellipsoid shape for the amplitude $s(\mathbf{r})$, whose implicit form is

$$\left( \left| \frac{x}{a} \right|^{2/e} + \left| \frac{y}{b} \right|^{2/e} \right)^{e/n} + \left| \frac{z}{c} \right|^{2/n} = 1, \tag{4}$$

where the exponents parameter $n$ and $e$ control the roundedness of the particle. $a$, $b$, and $c$ are the upper bounds of the particle size along the $x$, $y$, and $z$ directions, respectively. All of these values were selected from random distributions to create a diverse set of shapes. For the phase distribution $\phi(\mathbf{r})$ of the particles, a 3D Gaussian correlated profile[28,38] was used, which is given as

$$\phi(\mathbf{r}) = \frac{L_x^{1/2} L_y^{1/2} L_z^{1/2}}{\pi^{3/4}} \iiint e^{-\frac{(x-x')^2}{2L_x^2} - \frac{(y-y')^2}{2L_y^2} - \frac{(z-z')^2}{2L_z^2}} z_u(x,y,z) dx' dy' dz' , \tag{5}$$

where, $z_u(x,y,z)$ obeys an uncorrelated Gaussian random distribution. $L_x$, $L_y$ and $L_z$ are the transverse correlation lengths along the $x$, $y$, and $z$ directions, respectively. During the simulation of the 3D diffraction patterns, the phase of the simulated particle was scaled and shifted to [0, 1], and outside the particle, the phase is set to zero. The generated training dataset contains a wide variety of amplitude and phase states.

**Supervised Learning Approach**

The 3D CNN model was trained in a supervised approach on pairs of real-space objects and their reciprocal-space diffraction patterns. We used a loss function $l_s$, to constrain the real and reciprocal space data at the same time:

$$l_s = \frac{1}{\alpha_1 + \alpha_2 + \alpha_3} \left[ \alpha_1 L_1(A_p, A_g) + \alpha_2 L_2(\phi_p, \phi_g) + \alpha_3 L_3(\sqrt{I_p}, \sqrt{I_g}) \right], \tag{6}$$

where $L_1(x_p, x_g) = L_2(x_p, x_g) = \frac{\sqrt{\sum_n (x_p - x_g)^2}}{\sqrt{\sum_n x_g^2}}$ and $L_3(x_p, x_g) = 1 - \frac{\sum_n |x_p - \bar{x}_p| \cdot |x_g - \bar{x}_g|}{\sqrt{[\sum_n (x_p - \bar{x}_p)^2][\sum_n (x_g - \bar{x}_g)^2]}}$. In

Eq. (7), $L_1$ and $L_2$ are the loss function for the amplitude and phase of the particle in real space,

separately. $L_3$ is the loss function for the X-ray diffraction intensity in reciprocal space, which is used to constrain the relation between the predicted amplitude and phase from the ML model in reciprocal space. Here, the subscript p denotes the predicted result from ML model, and the subscript g denotes the corresponding ground truth. $L_3$, is based on the Pearson correlation coefficient. For the training, we used $\alpha_1 = 1$, $\alpha_2 = 1$ and $\alpha_3 = 1$. The proposed CNN model was

implemented based on the Pytorch platform using Python[39]. When training the CNN model, we

adopted two optimizers: Adaptive Moment Estimation (ADAM) and Stochastic Gradient Descent

(SGD) to optimize the weights and biases of the CNN model[40,41]. During the training, the two optimizers were switched every 25 epochs for a total of 100. The start learning rate for both optimizers were 0.01, and after every 25 epochs, the learning rate was reduced by a factor of 0.95.

In our study, the size of the input 3D coherent X-ray diffraction pattern was $64 \times 64 \times 64$ pixels.

The training was completed on a computer with 256 GB of RAM and two NVIDIA Quadro V100

GPUs.

**Unsupervised Learning Approach**

When the 3D CNN model was used in unsupervised learning approach, only the 3D coherent X-

ray diffraction pattern was available as input. During the optimization, the loss function $l_u$ was

defined as:

$$l_u = \frac{1}{\beta_1 + \beta_2} \left[ \beta_1 L_3(\sqrt{I_p}, \sqrt{I_m}) + \beta_2 L_4(\sqrt{I_p}, \sqrt{I_m}) \right], \tag{7}$$

where $I_m$ is the measured or validation 3D coherent X-ray diffraction intensity. $I_p = \left| FT \rho_p(\mathbf{r}) \right|^2$ is the calculated 3D diffraction intensity. $\rho_p(\mathbf{r})$ is the complex particle density predicted by the CNN model after zero-padding to the same size of input diffraction data $I_m$. $L_4$ is the conventional $\chi^2$ error function defined as $L_4 = \frac{\sum_n \left( \sqrt{I_p} - \sqrt{I_m} \right)^2}{\sum_n I_m}$. We used two different ways to initiate the CNN model's weight and bias parameters, either from our pre-trained CNN model (*i.e.,* transfer learning) or by using random numbers. In Eq. (7), the weighting coefficients $\beta_1$ followed a modified Weibull distribution:

$$\beta_1 = a_0 \frac{k}{\lambda} \left( \frac{n}{\lambda} \right)^{k-1} e^{-(n/\lambda)^k} + a_1,$$  (8)

where $k = 1$, $\lambda = 0.5$. $n$ is the training epoch. $a_0$ and $a_1$ are the scale factors to let $\beta_1$ gradually change from $10^4$ to 1 during the training, as shown in Supplementary Figure 5, while $\beta_2$ remained equal to 1. Two optimizers, ADAM and SGD were utilized to optimize the results, switching every 200 epochs. The learning rate for both optimizers started at 0.006 and after every 200 epochs the learning rate was reduced by a factor of 0.95. When the CNN model is applied to the experimental Bragg diffraction data, due to the existence of the shear distortion effects[42-44] in Bragg CDI, all the predicted results were converted from detector to laboratory coordinates after zero-padding to the same size of input diffraction data.

**Bragg CDI Experiments**

The Bragg CDI experiments were performed at 34-ID-C at Advanced Photon Source (APS), Argonne National Laboratory, USA. A front-end horizontal slit of 100 μm was used to improve the source coherence, and a double crystal monochromator was used to set the energy of the incident X-ray to 9 keV. A coherent beam of $30 \times 70$ μm$^2$ was selected and focused to ~

$630 \times 470$ nm$^2$ by Kirkpatrick-Baez (KB) mirrors before impinging on the samples. The four samples were chemically synthesized in nanocrystal format by different methods and attached to silicon wafer substrates for handling. The corresponding 3D coherent diffraction intensities were obtained by a rocking curve of the target Bragg peak of the samples, {101} for $BaTiO_3$ and $SrTiO_3$ and {111} for Au and Pd, as a series of 2D coherent diffraction patterns were recorded by a Medipix detector with $55 \times 55$ μm$^2$ pixels. In the figures, all experimental Bragg coherent X-ray diffraction patterns are presented in laboratory coordinates[42-44].

**Conventional Iterative Phase-Retrieval Method**

For the conventional iterative phase retrieval, the measured Bragg 3D diffraction patterns in detector coordinate were used as input to an iterative phase-retrieval scheme described by Robinson & Harder[7] to reconstruct their corresponding real-space particles' information, separately. During the reconstruction, the initial particle was obtained by inverse Fourier transformation of the amplitude of the input diffractions pattern with a random phase distribution, whose corresponding range is $[-\pi, \pi]$. The initial support size of the particle in real space is half the size of the input diffraction pattern array in each dimension. The algorithm starts with 50 steps of error reduction. Then, it was switched between hybrid input-output with $\beta = 0.9$ and error reduction after every 50 iterations. After 100 iterations, the shrink-wrap method[45] was applied in real space to dynamically update the support every ten iterations. At the end, 100 steps of error reduction were used to assure convergence. The total number of iterations was 2000. After reconstruction, all the reconstructed results were converted from detector to laboratory coordinates[42-44]. All the isosurfaces shown in the paper are plotted by using the open-source ParaView[46].

**DATA AVAILABILITY**

The source data that support the findings of this study are available upon request from the corresponding author.

## CODE AVAILABILITY

The python codes used in this study will be made available to readers upon request to the corresponding authors.

## ACKNOWLEDGEMENTS

Work at Brookhaven National Laboratory was supported by the U.S. Department of Energy, Office of Science, Office of Basic Energy Sciences, under Contract No. DE-SC0012704. J. D. received funding from the China Scholarship Council (CSC). Work at UCL was funded by EPSRC. Measurements were carried out at the Advanced Photon Source (APS) beamline 34-ID-C, which was supported by the U. S. Department of Energy, Office of Science, Office of Basic Energy Sciences, under Contract No. DE-AC02-06CH11357. The beamline 34-ID-C was built with U.S. National Science Foundation grant DMR-9724294.

## COMPETING INTERESTS

The authors declare no competing interests.

## AUTHOR CONTRIBUTIONS

L.W., S.Y., and I.K.R. developed the convolutional neural networks and performed the experimental data analysis. I.K.R., L.W., A.F.S., T.A.A., J.C., R.H., and W.C. carried out the BCDI experiments at 34-ID-C at Advanced Photon Source (APS), Argonne National Laboratory, USA. L.W. and I.K.R. wrote the manuscript and all the authors contributed to discussion of the manuscript.

## ADDITIONAL INFORMATION

Supplementary information is available for this paper at XX.

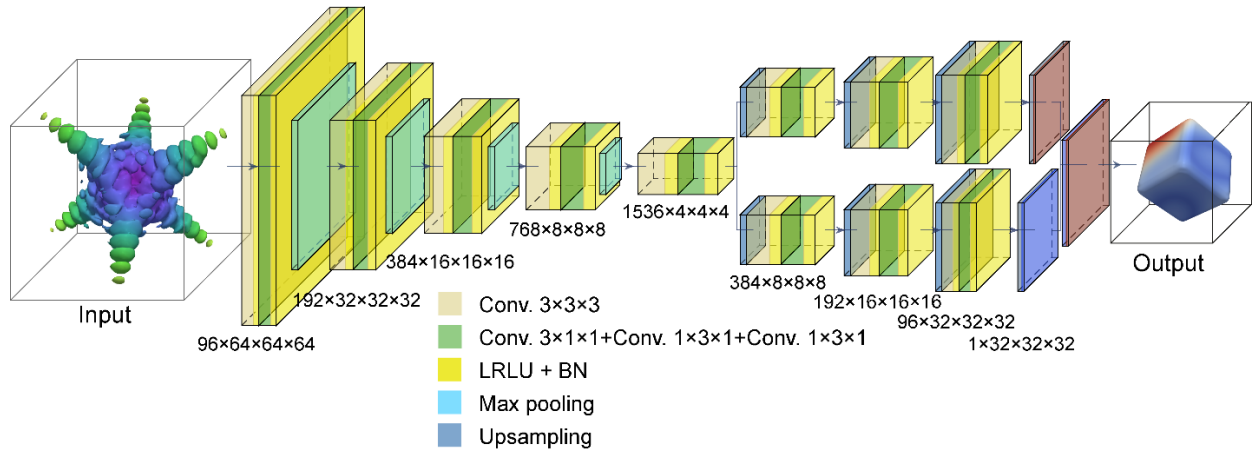

**Fig. 1 Overall scheme of 3D deep neural network for single-particle coherent diffraction imaging inversion.** The proposed 3D neural network is comprised of an encoder network and two decoder networks. In the network, the amplitude of a 3D coherent X-ray diffraction pattern in reciprocal space is used as input, and the output is the complex structure information (*i.e.,* amplitude and phase) of the particle in real space.

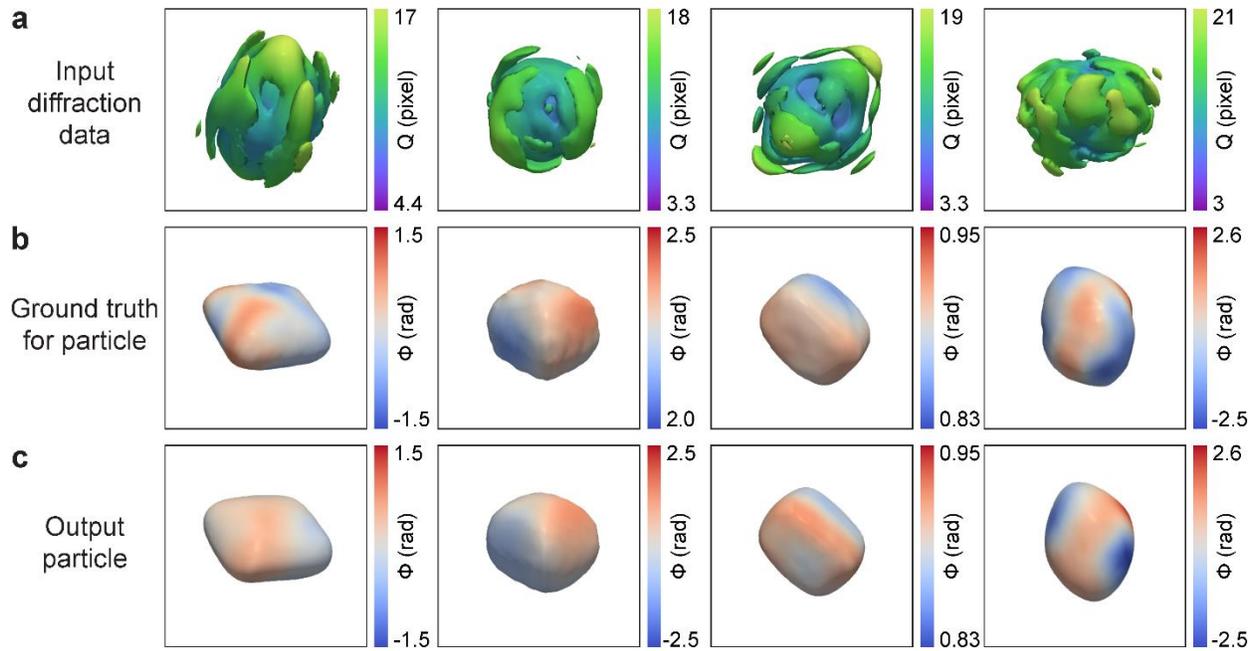

**Fig. 2 Performance of trained 3D CNN model in testing data. a** 3D isointensities of test input coherent X-ray diffraction patterns, which were not used for training. Here, the colors correspond to the radial distance from the origin of the reciprocal space. **b** Isointensity of the ground truth for the corresponding particles. **c** The complex-valued image predicted by the CNN model. Here, the isosurface plots in (**c**) and (**b**) are obtained by the amplitude of the particles and the corresponding color represents the phase distribution on their surfaces.

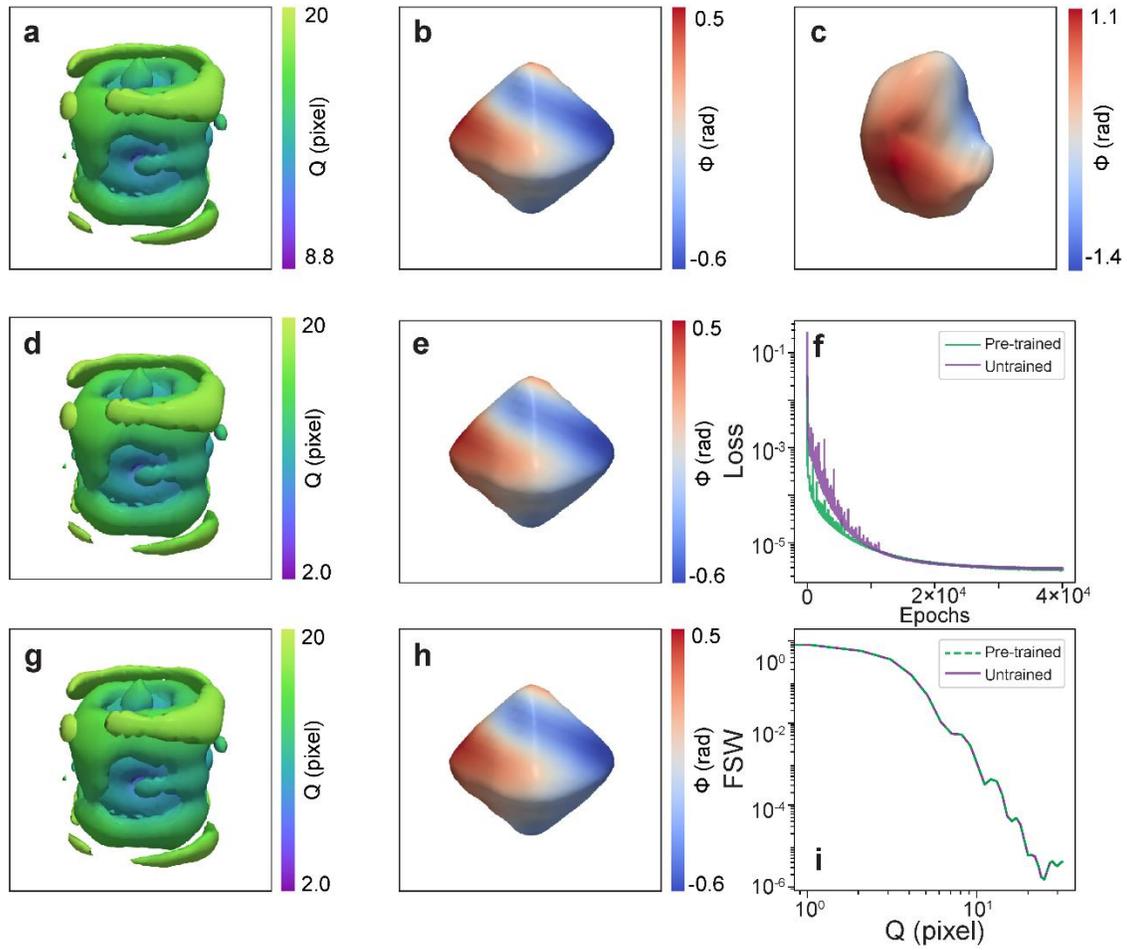

**Fig. 3 Representative results using 3D CNN model in unsupervised learning mode. a** Input coherent diffraction pattern and (**b**) corresponding ground truth of the particle. **c** Predicted result from the trained 3D CNN model. **d** Calculated coherent X-ray diffraction pattern. **e** Predicted particle using the pre-trained CNN model in the unsupervised transfer learning approach with all weights available for optimization. **f** Loss (or error) as a function of the training epochs for the CNN model during unsupervised transfer learning. (**g**)-(**h**) Same using the CNN model in the unsupervised learning approach with random initialization (*i.e.,* without transfer). **i** Fourier spectral weights of predicted results plotted as a function of momentum transfer radius.

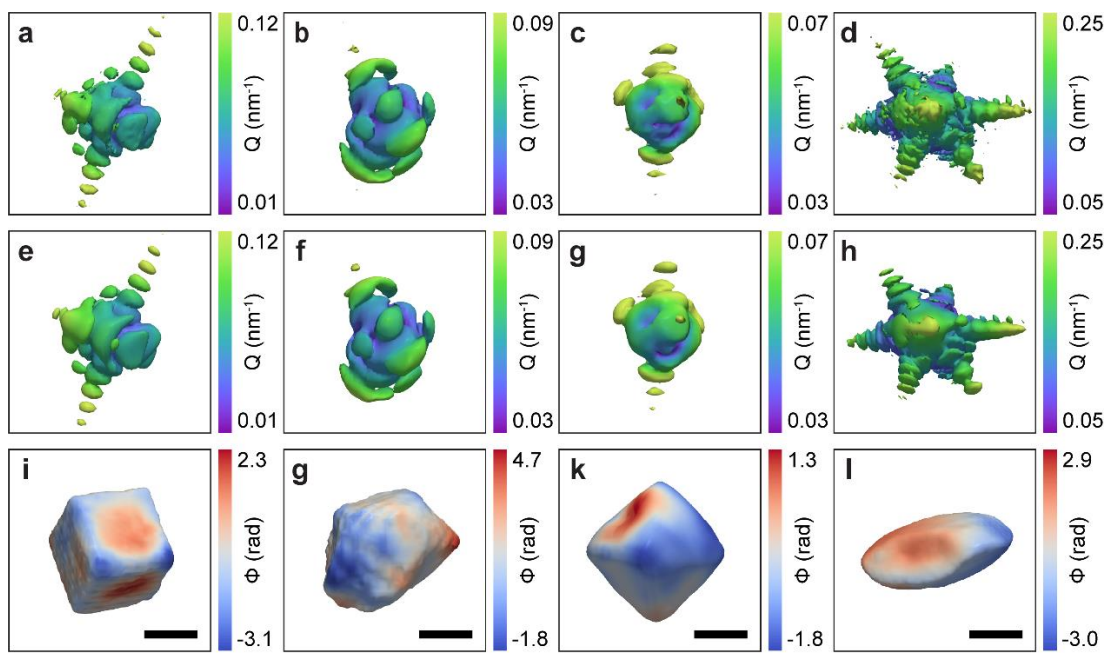

**Fig. 4 Performance of 3D CNN model on experimental coherent X-ray diffraction data.** (**a**)-(**d**) 3D plots of the isointensity for the measured 3D coherent X-ray diffraction patterns of $SrTiO_3$, $BaTiO_3$, Pd, and Au nanocrystals. (**e**)-(**h**) Corresponding isointensity plots of the 3D diffraction patterns of the predicted particles from the CNN model in the reciprocal space. The colors in (a)-(h) correspond to the radial distance from the origin of the reciprocal space. (**i**)-(**l**) The corresponding reconstructed real-space particle structures from the model. In (**i**)-(**l**) the surface colors encode the phase value on the surfaces of these particles. Here, the reconstructed results are presented in laboratory coordinates (see Methods for details). The scale bars are all 150 nm.

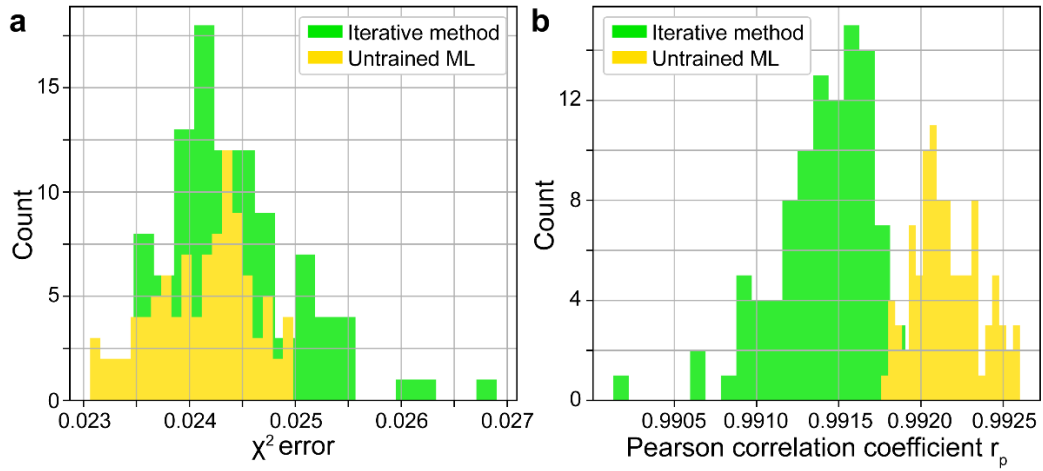

**Fig. 5 Comparison of error metrics for two different methods. a** Histogram of the observed $\chi^2$ for the reconstructions from the conventional iterative method and the CNN model starting from randomly initialized weight and bias parameters. **b** The corresponding histogram of the Pearson correlation coefficient for both methods.

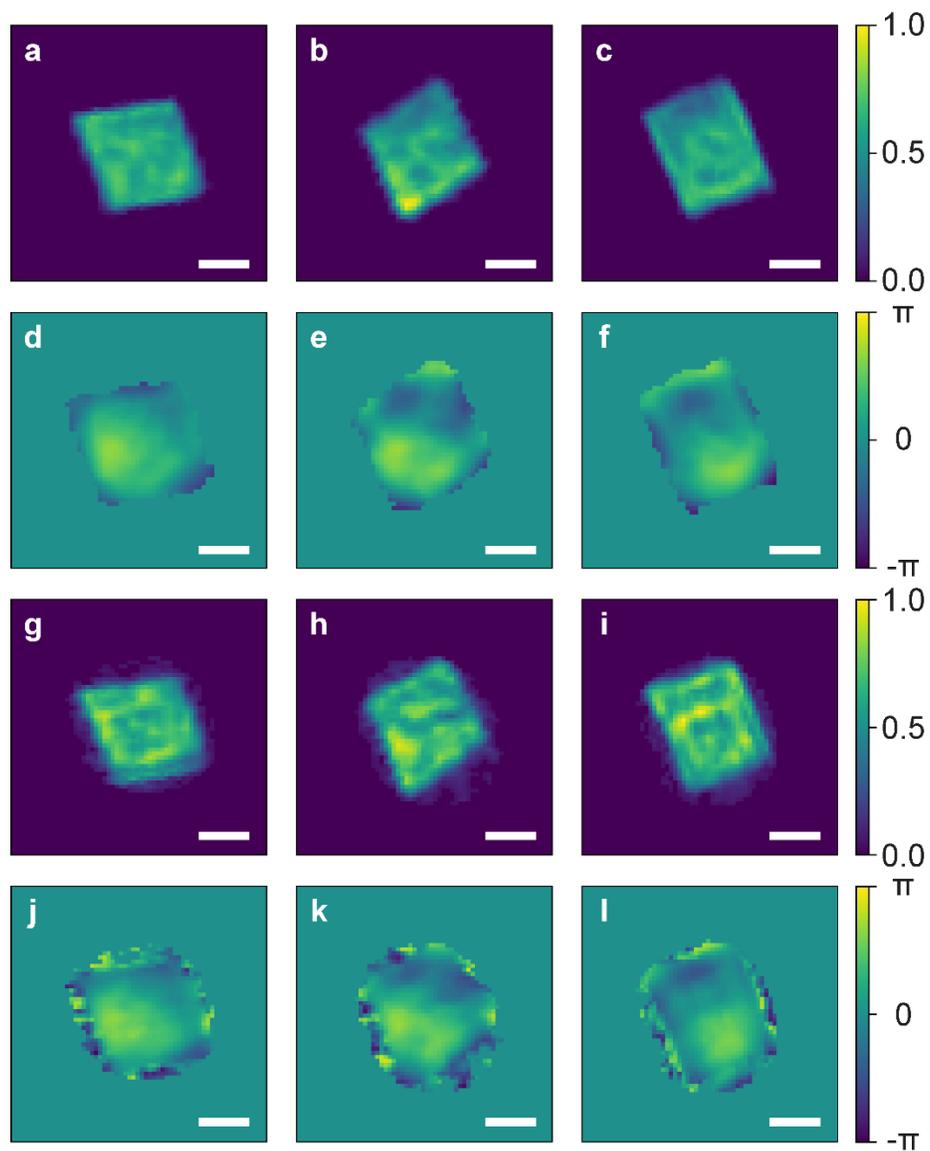

**Fig. 6 Comparison of experimental results for unsupervised learning and conventional iterative methods.** (**a**)-(**c**) Central X- Y- and Z-slices of the amplitude of the reconstructed SrTiO$_3$ particle, obtained with the CNN model. (**d**)-(**f**) The corresponding slices of the phase distribution. (**g**)-(**l**) Same for the iterative method. Here, the reconstructed results are presented in laboratory coordinates (see Methods for details). All the scale bars are 150 nm.